\documentclass[10pt,conference]{IEEEtran}

\usepackage{epsfig}
\usepackage{graphicx}
\usepackage{amsmath}
\usepackage{multirow}

\begin{document}


\title{ Bulk File Download Throughput in a Single Station WLAN with Nonzero Propagation Delay }

\author{Pradeepa BK and Joy Kuri \\ Centre for Electronics Design and Technology, \\
Indian Institute of Science, Bangalore. India. \\
{bpradeep,kuri}@cedt.iisc.ernet.in
}

\markboth{}
{ \MakeLowercase{}}

\maketitle

\begin{abstract}
We analyze TCP-controlled bulk file transfers in a single station (STA) WLAN 
with nonzero propagation delay between the file server and the WLAN. Our 
approach is to model the flow of packets as a closed queueing network (BCMP
network) with 3 service centres, one each for the Access Point (AP) and the 
STA, and the third for the propagation delay. The service rates of the first
two are obtained by analyzing the WLAN MAC. Simulations show a very close 
match with the theory.
\end{abstract}

\begin{IEEEkeywords}
WLAN, Access Points, TCP, Closed Queueing Network, BCMP Network.
\end{IEEEkeywords}

\IEEEpeerreviewmaketitle

\section{Introduction}\label{sec:Introduction}
We consider an infrastructure wireless network employing the IEEE 802.11 DCF mechanism and carrying TCP-controlled file downloads. In this 
paper, we study analytical models for evaluating  the performance of 
TCP-controlled downloads with non zero round trip propagation delays.
In the literature, many works address the analysis of TCP flows in 
wireless networks. However very few consider non negligible round trip propagation delays 
in their models. 

Our approach can be summarized as follows. First, we consider only two nodes
in the network -an AP and an STA. We obtain expressions for AP and STA throughput in 
the absence of propagation delays. We consider that all the wireless medium specific proprieties of the system are encapsulated in these expressions. Next, we include the effect of RTPD by using a closed queueing network. We validate our model by comparison with various simulated quantities, such as mean number of packets in the AP, STA and in flight. Our numerical results of model matches with maximum error of 1.87 \% error.

Outline of the paper: Section \ref{sec:Related_Work} outlines
related work. In Section \ref{sec:System_Model} we 
We summarize  out modelling assumption with the system model. In Section
\ref{sec:Analysis} we provide analysis for throughputs in case of two 
contending nodes. In Section \ref{sec:Evaluation}, we validate model 
by comparing with the results obtained from the Qualnet network simulator.
 In Section \ref{sec:Conclusion} we discuss the results and conclude the paper.
\section{Related Work}\label{sec:Related_Work} 
Numerous models and analyses have been proposed for wireless networks with 
TCP-controlled traffic, but very few consider propagation delays. In 
\cite{astn_model:Leith}, RTT is considered for modelling the TCP traffic in
WLAN, but the authors' interest was in providing analysis supporting the 
service differentiation feature in 802.11e. The authors ensured fairness among
TCP data packets and TCP ACKs by utilizing the different Access Categories in
802.11e, and their analysis exploited this fairness to show that service 
differentiation is possible.

A detailed analysis and modelling of the aggregate throughput of TCP flows in
WLANs for a single rate access point is given in \cite{astn_model:Kuriakose} 
and \cite{astn_model:Bruno4} by assuming negligible or zero round trip time 
(RTT). Similarly, the performance of the AP is evaluated in the multi rate 
case in \cite{astn_model:Krusheel}, \cite{astn_model:pradeep_kuri} and 
\cite{astn_model:pradeep_kuri2}. However these works also ignore the RTT.

An extension of this model in \cite{astn_model:Krusheel} considers two rates 
of association with long file uploads from STAs to a local server. The 
multirate case is considered in \cite{astn_model:pradeep_kuri}, and arbitrary
TCP windows are allowed in \cite{astn_model:pradeep_kuri2}. 
\cite{astn_model:Bruno1} and \cite{astn_model:Bruno2} present another analysis
 of a scenario of TCP-controlled upload and download file transfers, with UDP
traffic. They ignored the RTT effect on the behaviour of the network. The 
letter \cite{astn_model:Bruno3} gives the average value analysis of TCP 
performance with upload and download traffic with out considering RTT.
In \cite{astn_model:Onkar}, a finite buffer AP with TCP traffic in both upload 
and download directions is analysed with delayed and undelayed ACK cases. They 
consider server system located on the Ethernet to which the AP is connected 
and hence number of packets ``in flight'' outside the WLAN is ignored. 

\cite{astn_model:duffy} and \cite{astn_model:sudarev} provide models in finite 
load conditions by approximating the packet arrival precess at the STAs as
a Poisson process. \cite{astn_model:duffy} extended saturated model proposed in
\cite{astn_model:bianchi} to non saturated model by introducing a term for 
probability of queue being empty.  \cite{astn_model:sudarev} models every STA 
with M/G/1 queue considers the propagation delay as the delay in transmission
of packet in wireless medium, and not the time spent in reaching the WLAN
from the server. 

\cite{astn_model:Yu} provides an analysis for a given number of STAs and 
maximum TCP receive window size by using the well known $p$ persistent model 
proposed in \cite{astn_model:Cali}. However, both \cite{astn_model:Yu} and
\cite{astn_model:Cali} consider no RTT effect in the traffic.

\cite{astn_model:Miorandi} considers HTTP traffic. A queuing 
model is proposed to compute the mean session delay in the presence of 
short-lived TCP flows and the impact of TCP maximum congestion 
window size on this delay is studied. The analysis is extended to consider 
delayed ACKs asw well.
\section{System Model}\label{sec:System_Model}
\begin{figure}
\centering
\includegraphics[scale=0.5]{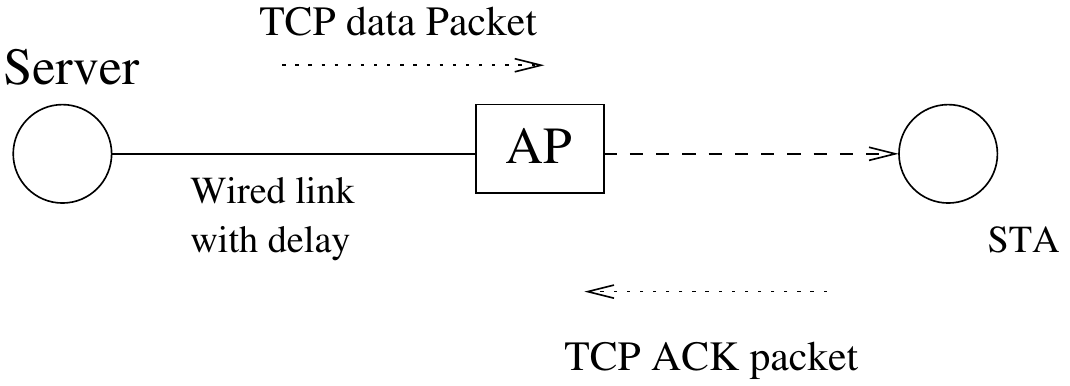} 
\caption{The network and traffic configurations. An STA is downloading 
a long file from a server through an AP.}
\label{fig:AP_1STA}
\end{figure}

We consider a WLAN in which an STA is associated with an AP as shown
in Figure \ref{fig:AP_1STA}. We consider long-lived TCP connections having 
bulk data to send. That is, we focus on large file transfers. The server 
is far away from the LAN. Hence, there is propagation delay between the AP 
and the server. Every packet experiences this delay. Figure \ref{fig:AP_1STA}
shows the direction of TCP flow. The STA has a single TCP connection to 
download long files from the server. Further, because of bulk transfer 
scenario, we can assume that the TCP flow is in the steady state. We assume
that the application at the receiver reads data from the socket receive buffer
at the rate at which it is received from the network. Hence, the TCP ACK 
packets always advertise the maximal TCP receive window size. Therefore, TCP 
startup transients can be ignored. Also, TCP timeouts do not occur. 

The AP and the STA contend for the channel using the DCF mechanism.
We assume that there are no link errors. Packets in the medium are lost only 
due to collisions between transmission attempts of AP and STA.
Further, we assume that all the nodes use the RTS-CTS mechanism while sending
packets, and use basic access to send ACK packets. As soon as the station 
receives a data packet, it generates an ACK packet without any delay and it 
is enqueued at the MAC layer for transmission. We assume that all the nodes 
have sufficiently large buffers, so that packets are not lost due to buffer
overflow. 
\section{Analysis}\label{sec:Analysis}
Because two entities, i.e., the AP and the STA, are contending for the channel,
we can consider probability of attempt as $ \beta _2 $, as in 
\cite{astn_model:kumar}. This amounts to assuming that both AP ans STA are 
backlogged permanently, i.e., that saturation conditions apply. Current TCP 
implementations use very large windows (window scaling option), and with these,
our assumption is justifiable.

\begin{figure}
\centering
\includegraphics[scale=0.5]{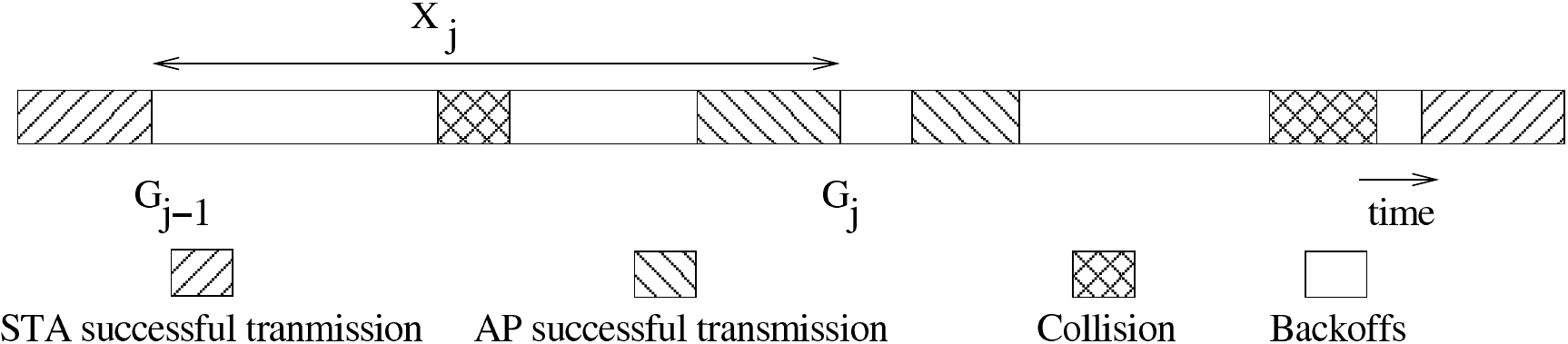} 
\caption{ An example of activities in a channel.Here, $G_j$ are random time
 instants at which successful transmissions complete for a packet. Random 
 duration $X_k$ denotes the $j^{th}$ contention cycle $[G_{j-1}, G_{j}) $. 
 Each contention cycle consists of one or more back off periods and collisions 
 but ends with a successful transmission. }
\label{fig:channel_activity}
\end{figure}

Figure \ref{fig:channel_activity} shows one possible sample path of events on 
the wireless channel in the WLAN with an AP and an STA. The random epochs 
$ G_j $ indicate the end of the $ j^{th} $ successful transmission from either
the AP or the STA.

Let $ T_sAP $ be the time taken by the AP to transmit one packet TCP packet 
including MAC and PHY headers. This can be obtained by using parameter 
shown in Table \ref{tab:parameter}. We have  
\begin{equation}
\begin{split}
TsAP & =  T_p + T_{PHY} + \frac{ L_{RTS} }{ r_c } + T_{SIFS} \\
 & + T_p + T_{PHY} + \frac{L_{CTS} }{ r_c } + T_{SIFS} + T_p + T_{PHY} \\
 & + \frac{L_{MAC} + L_{IPH} + L_{TCPH} + L_{TCP} }{ r_d } + T_{SIFS} \\
 & + T_p + T_{PHY} + \frac{ L_{ACK} }{ r_c } + T_{DIFS} 
\end{split}
\label{eq:tsAP}
\end{equation}
As  data transmission follows an RTS-CTS exchange, the lengths of the RTS and
CTS packets, as well as $T_{SIFS}$, are used in \eqref{eq:tsAP}
\begin{table}
\begin{tabular}{|p{3.0cm}|p{2.0cm}|p{2.0cm}|}
\hline Parameter &Symbol &Value \\ \hline
 PHY data rate& $r_d$& 11 Mbps  \\ \hline
 Control rate& $r_c$& 2 Mbps  \\ \hline
 PLCP preamble time& $T_p$ & $144   \mu s$ \\ \hline
 PHY Header time& $T_{PHY}$ & $48   \mu s$   \\ \hline
 MAC Header size& $L_{MAC}$& 34 bytes \\ \hline
 RTS Header size& $L_{RTS}$& 20 bytes \\ \hline
 CTS Header size& $L_{CTS}$& 14 bytes \\ \hline
 MAC ACK Header size& $L_{ACK}$& 14 bytes \\ \hline
 IP Header size& $L_{IPH}$& 20 bytes \\ \hline
 TCP Header size& $L_{TCPH}$& 20 bytes \\ \hline
 TCP ACK Packet size & $L_{TCP-ACK}$& 20 bytes \\ \hline
 TCP data payload size &$L_{TCP}$& 1460 bytes \\ \hline
 System slot time&$\delta$&   $20 \mu s$  \\ \hline
 DIFS time &$T_{DIFS}$& $50 \mu s$   \\ \hline
 SIFS time &$T_{SIFS}$& $10 \mu s$   \\ \hline
 EIFS time &$T_{EIFS}$& $364 \mu s$   \\ \hline
 Min. Contention Window & $ CW_{min} $	&  31 \\ \hline
 Max. Contention Window & $ CW_{max} $ & 1023 \\ \hline
\end{tabular}
\caption{Values of Parameters used in Analysis and Simulation} \label{tab:parameter}
\end{table} 
We note that whenever there is a collision between RTS packet from the AP and a
TCP-ACK packet from the STA, the channel time wasted is that due to the TCP-ACK
packet. This is because the RTS packet is smaller than a TCP-ACK packet as 
given in \ref{tab:parameter}.  Let $T_c$ be the time spent in collision.
\begin{equation}
\begin{split}
T_c = T_p + T_{PHY} + \frac{L_{MAC} + L_{IPH} + L_{TCP-ACK} + L_{TCP} }{ r_d }\\
+ T_{EIFS} 
\end{split}
\end{equation}
In the above expressions, we have considered that TCP data packets are larger
than the RTS threshold and hence the AP uses the RTS-CTS access mechanism. 
Since TCP-ACKs are smaller, the STA uses the basic access mechanism. Let $T_{sSTA} $
be the time taken by the STA to transmit one TCP ACK packet, including MAC and PHY
overhead. We have 
\begin{equation}
\begin{split}
T_{sSTA} 	& = T_p + T_{PHY} \\
		& + \frac{L_{MAC} + L_{IPH} + L_{TCPH} + L_{TCP-ACK} }{ r_d } \\
		& + T_{SIFS} + T_P + T_{PHY} + \frac{L_{ACK} }{ r_c } + T_{DIFS} 
\end{split}
\end{equation}

The probability that the AP wins the contention in the first attempt is
$ \beta _2 (1- \beta _2) $; let this be denoted as $ \alpha _{2} $.

Let the mean back-off length at $ i^{th} $ attempt be 
\begin{equation*}
 EB^{i} = \frac{ 2^{i} CW_{min} + 1 }{2}
\end{equation*}
Let $s_{sAP}^{k}$ be the mean time taken by AP to transmit a TCP data packet
after $k$ collisions. This can be obtained as
\begin{equation}
s_{AP} ^{k} = \sum _{i=1} ^{k+1} EB^{i} + k T_{c} + T_{sAP}
\end{equation}
which is mean weighted sum of conditional means. In the above Equation,
if the back-off window reaches $ CW_{max} $, 
it stays fixed at $ CW_{max} $. 

Also the probability of $ k$ times collisions and the success at $(k+1)^{th}$ is 
$ (1-\alpha _2 )^k \alpha _2 $. 

Hence the mean service time at the AP, with the STA is also contending for the 
channel is given by 
\begin{equation}
 s _{AP} = \sum _{i =1}^{\infty} s_{AP}^{k} ( 1 - \alpha _2)^k \alpha _2
\end{equation}

Similarly, the time taken by the STA to transmit a TCP-ACK packet after
$k$ collision is
\begin{equation}
s_{STA} ^{k} = \sum _{i=1} ^{k+1} EB^{i} + k T_{c} + T_{sSTA}
\end{equation}
and the mean service time at the STA is
\begin{equation}
 s _{STA} = \sum _{i =1}^{\infty} s_{STA}^{k} ( 1 - \alpha _2)^k \alpha _2
\end{equation}

\subsection{BCMP network}
Having obtained the service rate of the AP and STA, we can model the scenario 
shown in Figure \ref{fig:AP_1STA} as a BCMP closed queueing network with 
three service centers \cite{astn_model:bcmp}. The queues in this network 
representing the AP and STA are First Come First Served queues (FCFS), which 
are `Type 1' service centres in the terminology of \cite{astn_model:bcmp}. 
Similarly, the queue representing round trip propagation delay (RTPD) is an 
infinite server queue with deterministic service time, which is a `Type 3' 
service center.

Let us consider $ W $ packets in this network. Let $ w_1 $ packets to be at 
center 1; that is, $ w_1 $ among $ W $ packets are at the AP. Also, let 
$w_2$ out of $W$ packets be at center 2, which is an STA. Similarly, the 
remaining packets, say $w_3$, are in the delay center. The state of the network
can be represented by $ S = (x_1, x_2, x_3) $,
as in \cite{astn_model:bcmp}. The definitions of $ x_1, x_2$ and $ x_3$ 
depend on the type of the service center  and are given in \cite{astn_model:bcmp}.

Every transition is both a departure from one center and an arrival at another 
center. For every $i$, let $ e_{i} $ be the fraction of transitions that are arrivals at
(departures from) center $i$. Let $ v_{i',i}$  be the probability that a customer at center
 $i'$  goes to center $i$. We have $v_{1, 2} = v_{1, 2} = v_{1, 2} = 1 $. From \cite{astn_model:Wolff}, $ e_{i} $  unique solution
(that sums to 1) of the following system of equation
\begin{equation}
e_{i} = \sum_{i} (e_{i'}) v_{i',i}
\end{equation} 
Hence, $e_{i}= \frac{1}{3}$.  Every packet moves from one service center to another. 
\begin{figure}
\centering
\includegraphics[scale=0.5]{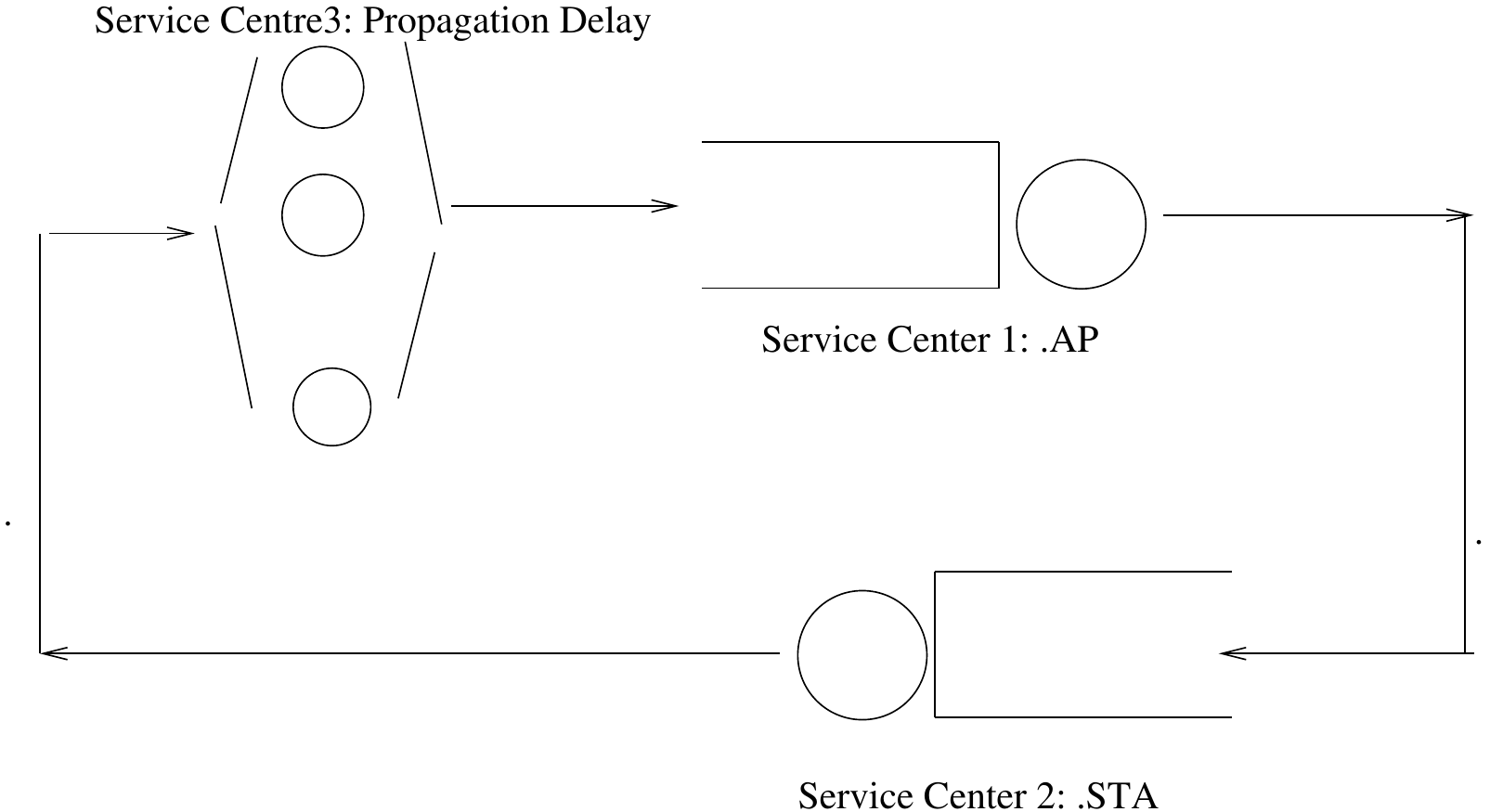} 
\caption{An equivalent BCMP closed queueing network model for the scenario given in 
Figure \ref{fig:AP_1STA} considering packets as customers. Total number of customers
is equal to the maximum receive window advertised by TCP receiver }
\label{fig:threeQs}
\end{figure}

By the BCMP theorem \cite{astn_model:bcmp}, the equilibrium probabilities are given by 
\begin{equation}
P(S = x_1,x_2, x_3) = C d(S)f_1 (x_1) f_2 (x_2) f_2 (x_3)
\label{eq:bcmp_dist}
\end{equation}
where $C$ is the normalizing constant chosen to make the equilibrium state
probabilities sum to 1. $d(S)$ is a function of the number of customers in 
the system, and $f_i$ is a function that depends on the type of service 
center $i$.

From \cite{astn_model:bcmp}, for the FCFS server, i.e., AP, center 1,
\begin{equation}
f_1(x_1) =  \left( s_{AP} \right) ^{n_1} e_{1} 
\end{equation}

for the FCFS server STA, center 2,
\begin{equation}
f_2(x_2) =  \left( s_{STA} \right) ^{n_2} e_{2}
\end{equation}

and for the infinite server, delay model, center 3,
is represented by cascading of $c$ number of exponential servers in $c$ stages
with service rate $ \frac{1}{c \times t_{RTPD}} $  method in \cite{astn_model:Arnold}
(by considering large value of $c$) gives 
\begin{equation}
f_3(x_3) =  \Pi _{l=1}^{c} \left( \frac{ e_{3}}{ c \times \tau_{RTPD} } \right) ^{n_{3,l} } (1/n_{3,l}!)
\end{equation}
For a closed network, $ d(S) = 1 $.
 
The average number of  packets at the AP,  $ n_{AP } $, the average number of packets
at the STA,  $ n_{ STA} $, and the average number of packets in propagation, $ n_{RTPD}$
can be obtained by finding the marginal distributions from \eqref{eq:bcmp_dist}. 
 
 From Figure \ref{fig:threeQs} the total number of packets is distributed among the three
service centers.  
\begin{equation*}
 n_{AP}  + n_{STA} + n_{RTPD}  = W  
\end{equation*}
Let the throughput in the closed network of Figure \ref{fig:threeQs} be $t_H$.
Then, applying Little's Theorem to service center 3, we have

\begin{equation}
n_{RTPD} = t_H \times t_{RTPD}
\end{equation}

 \section{Evaluation}\label{sec:Evaluation}
In this section, we compare the numerical results obtained from our analysis with
those obtained from simulation using  the  the Qualnet 4.5 network simulator \cite{astn_model:Qualnet}.
The parameters used were taken from  the IEEE 802.11b standard  and are given in Table 
\ref{tab:parameter}. The error bars in the simulation curves denote 95\% 
confidence intervals.  We have taken the STA to be associated  at rate 11 Mbps, with
 control packet  transmission rate at 2 Mbps. RTPD is varied from 10ms to 90ms
in step of 10ms. TCP Receive window is taken as 70 packets.

In Table \ref{table:AP_buffer}, the average number of packets in the AP obtained from analysis and simulations are shown.
  
\begin{table}[ht]
\centering 
\begin{tabular}{|c|c|c|c|c|}
\hline
& Analysis & \multicolumn{3}{|c|}{Simulation} \\ \hline
RTPD(ms) & Packets &	Mean 	&	Max 	&	Min \\ 
\hline
10	&	34.75	&	34.77	&	37.40	&	32.14	\\
20	&	29.89	&	29.93	&	33.05	&	26.81	\\
30	&	31.52	&	31.58	&	33.47	&	29.66	\\
40	&	28.85	&	28.90	&	30.68	&	27.12	\\
50	&	26.92	&	26.97	&	28.79	&	25.14	\\
60	&	26.78	&	26.84	&	28.22	&	25.46	\\
70	&	23.84	&	23.89	&	25.19	&	22.59	\\
80	&	22.34	&	22.40	&	23.46	&	21.33	\\
90	&	20.51	&	20.56	&	21.45	&	19.67	\\
100	&	19.27	&	19.32	&	19.94	&	18.71	\\
\hline 
\end{tabular}
\caption{Number of packets in AP buffer at rate \textit{11Mbps} for different values of RTPD.  } 
\label{table:AP_buffer} 
\end{table}
\begin{table}[ht]
\centering 
\begin{tabular}{|c|c|c|c|c|}
\hline
& Analysis & \multicolumn{3}{|c|}{Simulation} \\ \hline
RTPD(ms) & Packets &	Mean 	&	Max 	&	Min \\ 
\hline
10	&	3.19		&	3.19		&	3.21		&	3.18	\\
20	&	6.37		&	6.361	&	6.39		&	6.33	\\
30	&	9.55		&	9.536	&	9.59		&	9.48	\\
40	&	12.74	&	12.723	&	12.78	&	12.66	\\
50	&	15.92	&	15.89	&	15.96	&	15.82	\\
60	&	19.06	&	19.02	&	19.12	&	18.92	\\
70	&	22.22	&	22.168	&	22.31	&	22.03	\\
80	&	25.44	&	25.381	&	25.52	&	25.25	\\
90	&	28.57	&	28.492	&	28.61	&	28.37	\\
100	&	31.65	&	31.557	&	31.71	&	31.4	\\
\hline 
\end{tabular}
\caption{Number of packets in ``in flight'' for different values of RTPD.  } 
\label{table:RTT_buffer} 
\end{table}
\begin{table}[ht]
\centering 
\begin{tabular}{|c|c|c|c|c|}
\hline
& Analysis & \multicolumn{3}{|c|}{Simulation} \\ \hline
RTPD(ms) & Packets &	Mean 	&	Max 	&	Min \\ 
\hline
10	&	32.07	&	33.40	&	36.8		&	30.00	\\
20	&	33.74	&	33.14	&	35.46	&	30.82	\\
30	&	28.93	&	27.95	&	29.88	&	26.03	\\
40	&	28.40	&	28.55	&	30.17	&	26.94	\\
50	&	27.16	&	27.99	&	29.99	&	25.98	\\
60	&	24.16	&	24.12	&	25.44	&	22.79	\\
70	&	23.94	&	23.36	&	24.69	&	22.03	\\
80	&	22.22	&	22.59	&	23.69	&	21.47	\\
90	&	20.93	&	20.21	&	21.42	&	18.99	\\
100	&	19.09	&	19.41	&	20.04	&	18.78	\\
\hline 
\end{tabular}
\caption{Number of packets in STAs buffer at rate \textit{11 Mbps}  
for different values of RTPD.} 
\label{table:STA_buffer} 
\end{table}
 
\begin{table}[ht]
\centering 
\begin{tabular}{|c|c|c|c|c|}
\hline
& \multicolumn{2}{|c|}{Analysis} & \multicolumn{2}{|c|}{Simulation} \\ \hline
RTPD(ms) & packets/s &	Mean  & Max &	Min \\
\hline
10	&	320.26	&	320.34	&	321.47	&	319.21	\\
20	&	319.54	&	319.75	&	320.97	&	318.54	\\
30	&	319.45	&	319.64	&	321		&	318.29	\\
40	&	319.38	&	319.57	&	321.2	&	317.95	\\
50	&	319.07	&	319.15	&	320.46	&	317.86	\\
60	&	318.85	&	319.11	&	320.55	&	317.68	\\
70	&	318.49	&	318.75	&	320.34	&	317.17	\\
80	&	318.41	&	318.56	&	319.81	&	317.33	\\
90	&	318.32	&	318.45	&	319.71	&	317.21	\\
100	&	318.3	&	318.53	&	320.01	&	317.06	\\
\hline 
\end{tabular}
\caption{Throughput of the AP with a single STA at rate 11 Mbps for
different values of RTPD. TCP window = 100 packets} 
\label{table:AP_STA_thpt} 
\end{table}
 
\begin{figure}[ht]
\centering
\includegraphics[scale=0.7]{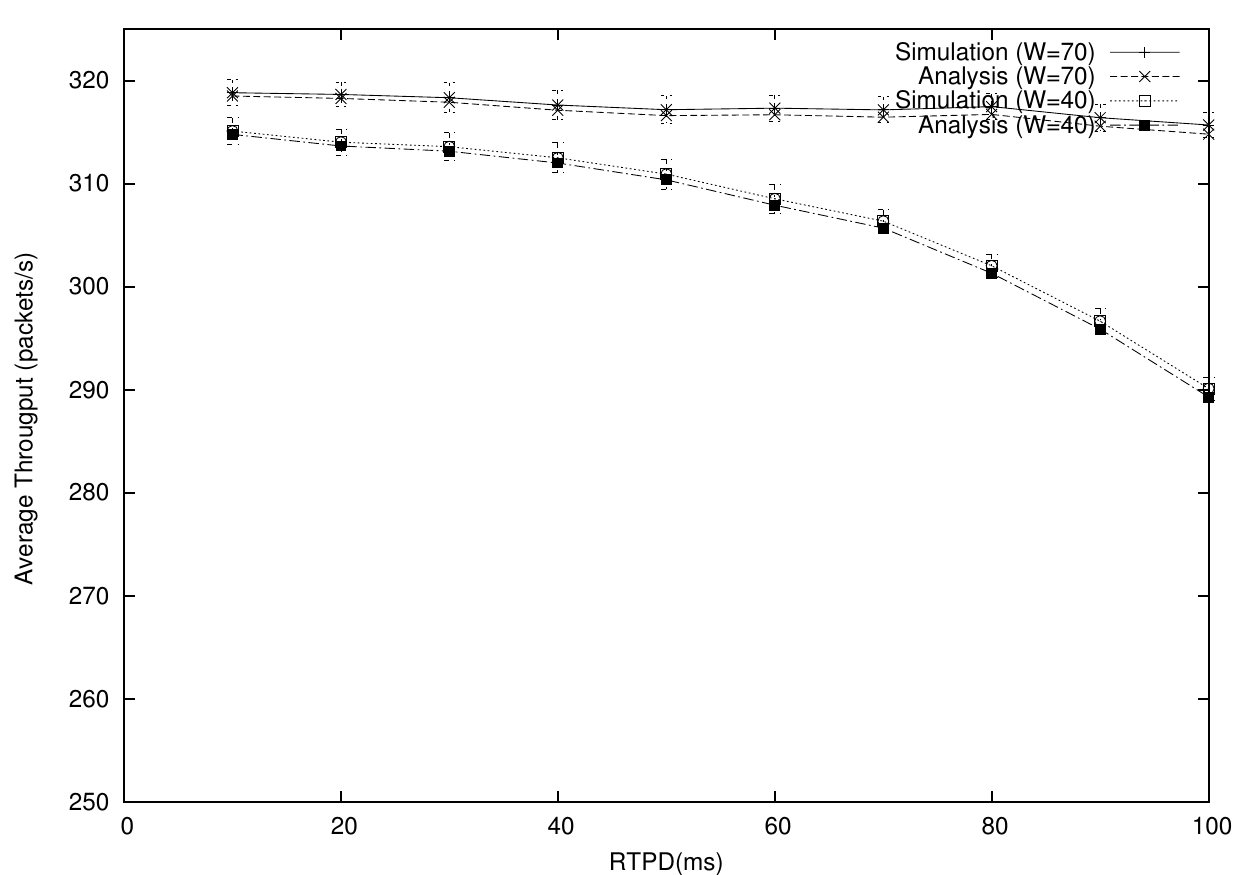} 
\caption{TCP throughput vs RTPD for TCP window 40 and 70 packets}
\label{fig:win_thpt1}
\end{figure} 

\begin{figure}[ht]
\centering
\includegraphics[scale=0.7]{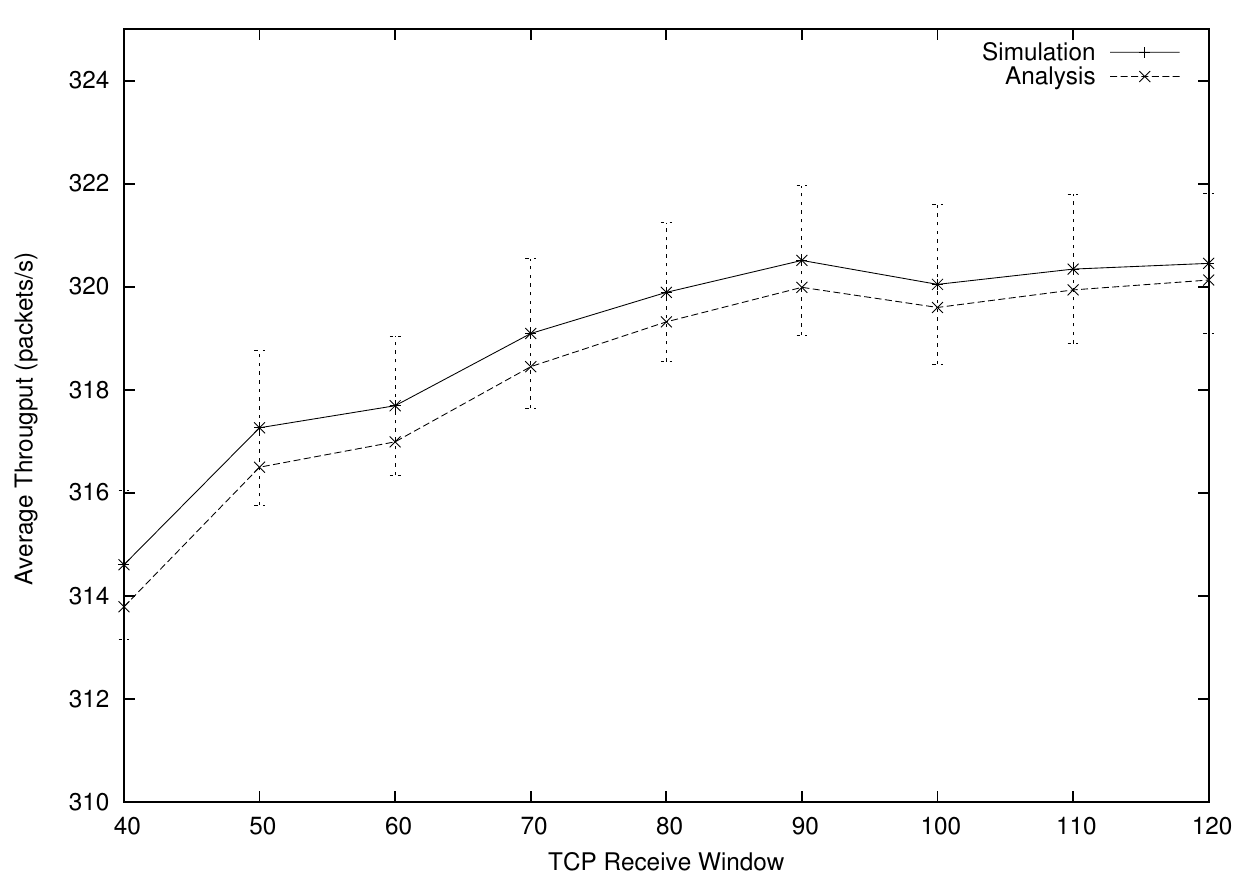} 
\caption{TCP throughput vs TCP window for RTPD of 10 ms}
\label{fig:thpt2}
\end{figure}

\begin{figure}[ht]
\centering
\includegraphics[scale=0.7]{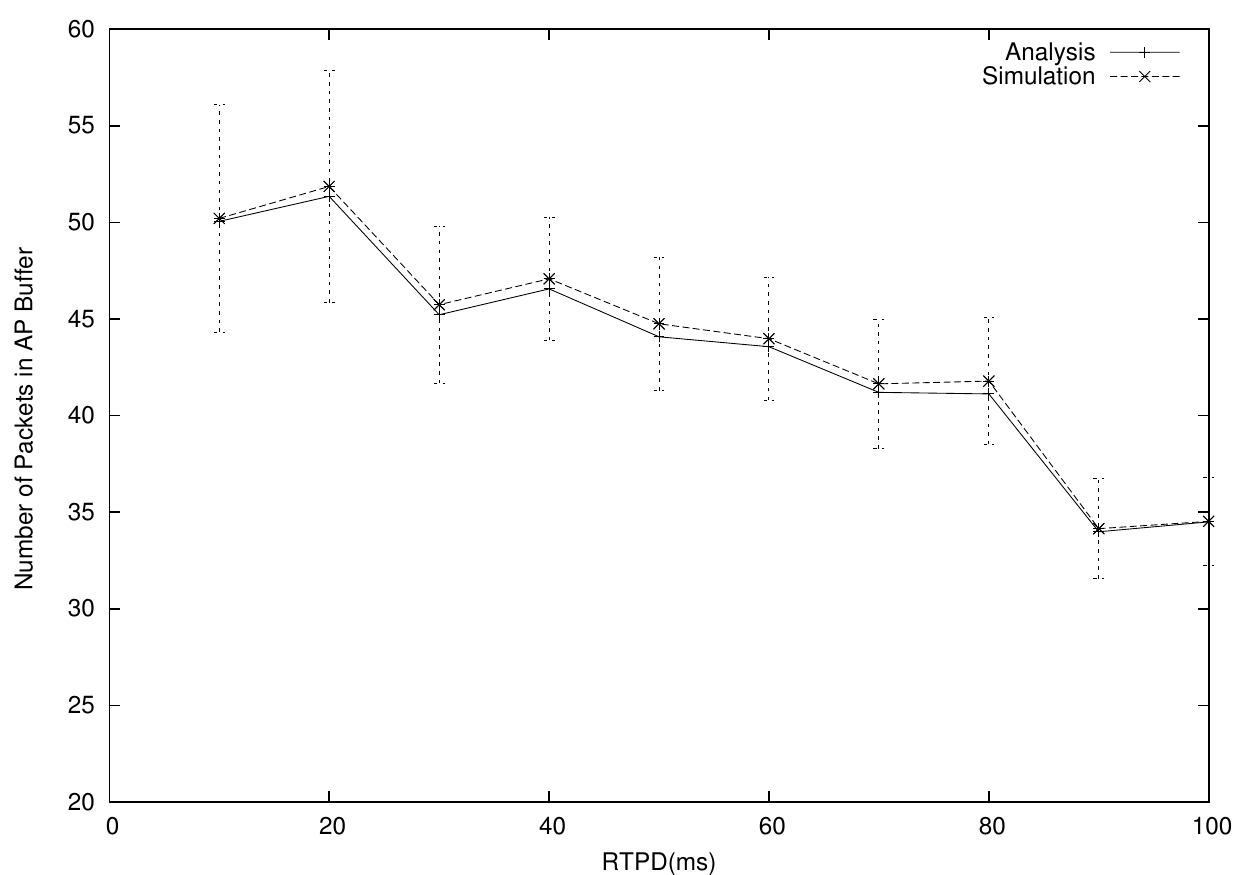} 
\caption{Number of Packets in AP buffer. TCP window = 100 packets.}
\label{fig:Nap}
\end{figure}

\begin{figure}[ht]
\centering
\includegraphics[scale=0.7]{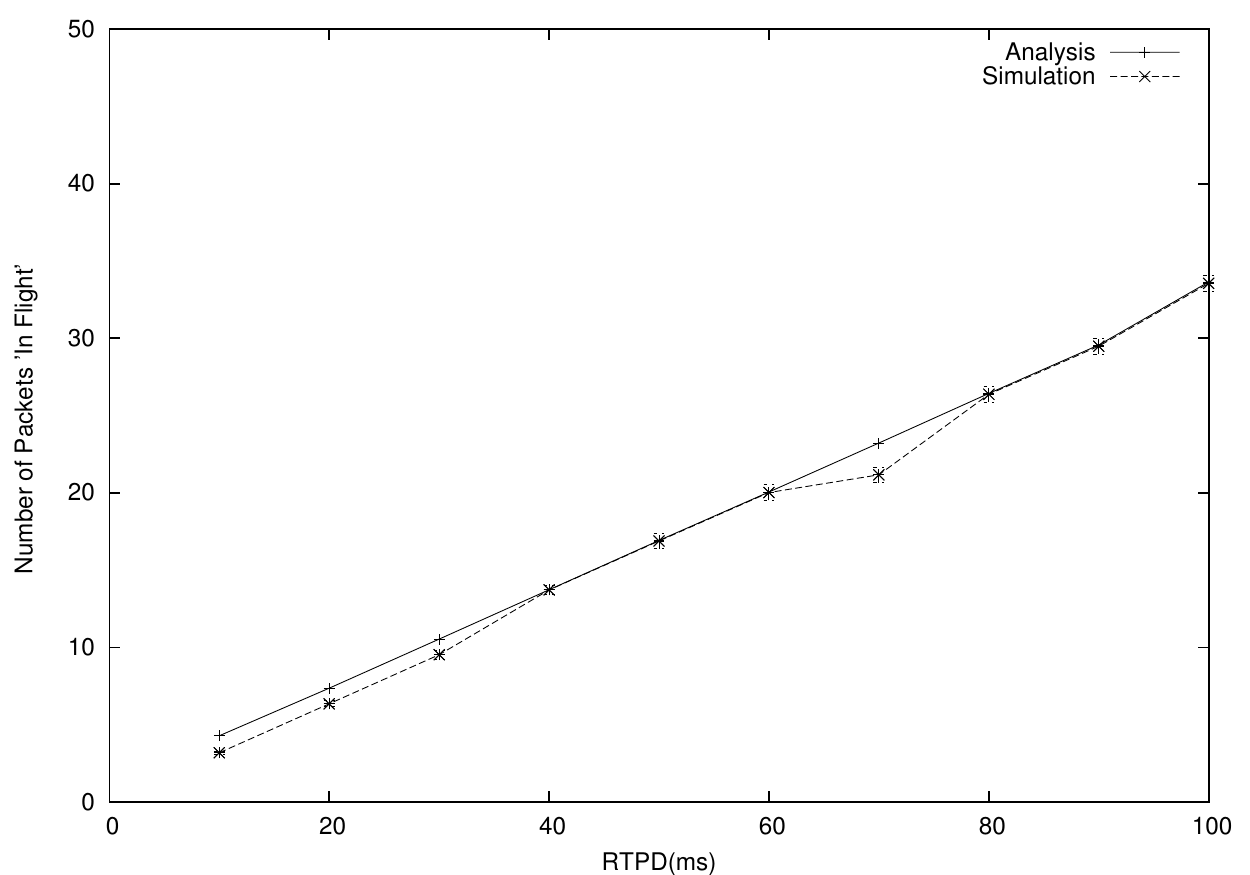} 
\caption{Number of Packets In Flight. TCP window = 100 packets.}
\label{fig:Npd}
\end{figure}

 In Tables \ref{table:AP_STA_thpt} and Figures \ref{fig:win_thpt1},
\ref{fig:thpt2}, \ref{fig:Nap} and \ref{fig:Npd} comparisons between analytical and simulation values are given for 11Mbps to illustrate the accuracy of the analytical model.
 
\section{Conclusion}\label{sec:Conclusion}
In this paper we developed a simple analytical framework to obtain accurate closed-form expressions for the performance of the AP and STAs with persistent TCP connections in the presence of round trip propagation delay. We verified the accuracy of the analytical model with the simulation results.We consider that TCP in its steady state, the TCP advertised window
is smaller than the TCP congestion window.





%


\ifCLASSOPTIONcaptionsoff
  \newpage
\fi

%




\end{document}